\documentclass[12pt]{article}
\usepackage{amsmath,amssymb}
\usepackage{graphicx}
\usepackage[pdftex]{hyperref}
\usepackage{caption}
\DeclareGraphicsExtensions{.eps,.bmp,.wmf,.jpg}
\numberwithin{equation}{section}
\def\be{\begin{equation}}
\def\ee{\end{equation}}

\def\bea{\begin{eqnarray}}
\def\eea{\end{eqnarray}}

\title{Dynamical analysis for a scalar-tensor model with kinetic and non-minimal couplings }
\author{L.N. Granda\thanks{luis.granda@correounivalle.edu.co} ,\, D. F. Jimenez\thanks{jimenez.diego@correounivalle.edu.co}\\{\it Departamento de Fisica, Universidad del Valle}\\{\it A.A. 25360, Cali, Colombia}}
\date{}
\begin{document}
\maketitle

\begin{abstract}
We study the autonomous  system for a scalar-tensor model of dark energy with non-minimal coupling to curvature and non-minimal kinetic coupling to the Einstein tensor. The critical points describe important stable asymptotic scenarios including quintessence, phantom and de Sitter attractor solutions. Two functional forms for the coupling functions and the scalar potential were considered: power-law and exponential functions of the scalar field. For power-law couplings, the restrictions on stable quintessence and phantom solutions lead to asymptotic freedom regime for the gravitational interaction. The model with dimensionless kinetic coupling constant gives stable de Sitter solutions. For the exponential functions the stable quintessence, phantom or de Sitter solutions, allow  asymptotic behaviors where the effective Newtonian coupling can reach either the asymptotic freedom regime or constant value. The phantom solutions could be realized without appealing to ghost degrees of freedom. Transient inflationary and radiation dominated phases can also be described.
\begin{description}
\item[PACS numbers]
98.80.-k, 95.36.+x, 04.50.Kd
\end{description}
\end{abstract}

\maketitle


\section{\label{intro}introduction}

The discovery of the current accelerated expansion of the universe,  confirmed by different observations  \cite{riess}, \cite{perlmutter}, \cite{kowalski}, \cite{hicken}, \cite{komatsu}, \cite{percival}, \cite{planck1}, \cite{planck2}, supposes a great challenge to understand the past history and the future destiny of our universe and the connection between the different stages of the cosmic evolution. This cosmological puzzle has been subject of intense investigations for almost the last two decades (see  \cite{copeland}-\cite{nojiriod} for review), and given the known problems with the cosmological constant \cite{starovinsky1}, \cite{peebles}, \cite{padmanabhan}, one of the promising sources to explain the nature of this accelerated expansion (the source of this expansion is called dark energy) is provided by the scalar-tensor models, which arise in different contexts. The models with general scalar field couplings to the curvature tensors like the non-minimal coupling to the curvature scalar or the scalar field derivatives coupled to the curvature tensors, as the models discussed here, fall within the general category of Horndeski theories \cite{horndeski}.
The non-minimal coupling between the scalar field and curvature appear in the process of quantization of the scalar field on curved space time \cite{ford, birrel, parker}, after compactification of higher dimensional gravity theories \cite{lamendola1} and in the context of string theories  \cite{metsaev}, \cite{meissner}. The non-minimally coupled scalar field applied to late time cosmology has been proposed by many authors to to address the dark energy problem since these couplings provide in principle a mechanism to evade the coincidence problem, allow phantom crossing in some cases \cite{perivolaropoulos},\cite{fujii}. The most studied coupling, the $F(\phi)R$ coupling, has been considered in different aspects, among others the constraint on the coupling by solar system experiments \cite{chiba}, the existence and stability of cosmological scaling solutions \cite{uzan, lamendola}, perturbative aspects and incidence on CMB \cite{perrotta, riazuelo}, tracker solutions \cite{perrotta1}, observational constraints and reconstruction \cite{boisseau, polarski, capozziello3} the coincidence problem \cite{tchiba}, super acceleration and phantom behavior \cite{vfaraoni,elizalde,carvalho,polarski1,nojiri3}, asymptotic de Sitter attractors \cite{vfaraoni1}, and a dynamical system for non-minimally coupled scalar field with power-law potential was studied in \cite{msami}.\\
Another important interaction is given by the couplings between the scalar kinetic term and curvature terms, which among others,  appear in the $\alpha'$-expansion of the string effective action \cite{metsaev, meissner}. These non-minimal derivative couplings to curvature were proposed in \cite{amendola, scapozziello} to study inflationary attractor solutions, in \cite{scapozziello1} to find a connection with the cosmological constant, in \cite{daniel} a kinetic coupling to the Ricci tensor was considered to study restrictions on the coupling parameter, and exact cosmological solutions and its asymptotical behavior were studied in \cite{vsushkov}. 
The non-minimal coupling of the scalar kinetic term to curvature leads to a rich variety of solutions for the different cosmological epochs, particularly for late time acceleration, as shown in \cite{sushkov2}-\cite{sushkov4}, \cite{gubitosi}, and a numerical study of general Horndeski type models applied to different physical properties of the universe have been performed in \cite{ebellini, zumalacarregui, rinaldi}. A non-minimal kinetic coupling to curvature with dimensionless coupling constant was considered in \cite{granda1}, where solutions with accelerated expansion were found, and in {\cite{granda2} a generalization of the model with an arbitrary coupling function of the scalar field was proposed and different late time cosmological solutions were studied. If in addition to the kinetic coupling we consider the Gauss-Bonnet curvature invariant coupled to the scalar field, then the resulting model gives us new viable cosmological solutions and enrich the dynamical behavior of the model as shown in \cite{granda5}-\cite{granda7}.\\
Despite the lack of sufficient astrophysical data to decide by one or another model, is still important to consider scalar tensor couplings to study late time Universe, since it could provide clues about how fundamental theories at high energies manifest at cosmological scales. In the present paper we study the late time cosmological dynamics for the scalar-tensor model with non-minimal and kinetic couplings to curvature. To this end, and due to the non-linear character of the cosmological equations, we use the dynamical system method to analyze the dynamics of the model and to derive asymptotic solutions that describe different cosmological scenarios with their stability properties.
The paper is organized as follows. In section II we introduce the model and give the general equations expanded on the FRW metric and introduce the dynamical variables. In section III we solve the equations for the critical points and give an analysis of the different solutions. In section IV we give a summary and discussion.

\section{The action and field equations}
The action for the scalar field and matter is given in equation (\ref{eq1}). The non-linear character of the cosmological equations  makes the integration of the same ones very difficult for a given set of initial conditions. Nevertheless the autonomous system for this model allows to study some interesting scaling solutions and the cosmological implications coming out from the different critical points.
\begin{eqnarray}\label{eq1}
S_{\phi}=&&\int d^{4}x\sqrt{-g}\Bigg[\frac{1}{2}F(\phi)R-\frac{1}{2}\partial_{\mu}\phi\partial^{\mu}\phi^{2}\nonumber\\
&&-V(\phi)+F_1(\phi)G_{\mu\nu}\partial^{\mu}\phi\partial^{\nu}\phi+{\cal{L}}_m\Bigg],
\end{eqnarray}
where
\be\label{eq1a}
F(\phi)=\frac{1}{\kappa^2}-h(\phi),
\ee
${\kappa}^2 =8\pi G$, $G_{\mu\nu}=R_{\mu\nu}-\frac{1}{2}g_{\mu\nu}R$, ${\cal{L}}_m$ is the Lagrangian for perfect fluid with energy density ${\rho _m}$ and pressure $ {P_m}$, $h(\phi)$ and $F_1(\phi)$ are the non-minimal coupling and kinetic coupling functions respectively. Note that the coefficient of the scalar curvature $R$ can be associated with an effective Newtonian coupling as $\kappa_{eff}^2=F(\phi)^{-1}$. Specializing to a spatially flat Friedmann-Robertson-Walker (FRW) metric
\begin{equation}\label{eq2}
ds^2=-dt^2+a(t)^ 2\sum_{i=1}^{3}(dx_i)^2,
\end{equation}
the Friedmann equations with Hubble parameter $H=\dot{a}/a$ can be written as follows
\be\label{eq3}
3H^2(F-3F_1\dot{\phi}^2)=\frac{1}{2}\dot{\phi}^2+V-3H\dot{F}+\rho_m
\ee
\be\label{eq4}
2\dot{H}(F-F_1\dot{\phi}^2)=-\dot{\phi}^2-\ddot{F}+H\dot{F}-6H^2F_1\dot{\phi}^2+2H\frac{d}{dt}(F_1\dot{\phi}^2)
-(1+w_m)\rho_m
\ee
\be\label{eq5}
\begin{aligned}
\ddot{\phi}+3H\dot{\phi}+&\frac{dV}{d\phi}-3(2H^2+\dot{H})\frac{dF}{d\phi}+3H^2\left(2F_1\ddot{\phi}+\frac{dF_1}{d\phi}\dot{\phi}^2\right)\\
& +6HF_1\left(3H^2+2\dot{H}\right)\dot{\phi}=0
\end{aligned}
\ee
The Eq. (\ref{eq1}) can be rewritten as
\be\label{eq6}
1-\frac{3F_1\dot{\phi}^2}{F}=\frac{\dot{\phi^2}}{6H^2F}+\frac{V}{3H^2F}-\frac{\dot{F}}{HF}+\frac{\rho_m}{3H^2F}
\ee
which allows us to define the following dynamical variables
\be\label{eq7}
\begin{aligned}
&&x=\frac{\dot{\phi}^2}{6H^2F},\;\;\; y=\frac{V}{3H^2F},\;\;\; f=\frac{\dot{F}}{HF}\\
&& k=\frac{3F_1\dot{\phi}^2}{F},\;\;\; \Omega_m=\frac{\rho_m}{3H^2F},\;\;\; \epsilon=\frac{\dot{H}}{H^2}
\end{aligned}
\ee
In terms of the variables (\ref{eq7}) the Friedmann equation (\ref{eq3}) becomes the restriction
\be\label{eq8}
1=x+y-f+k+\Omega_m
\ee
Due to the interaction term in the denominator, the density parameters $\Omega_m$ and $\Omega_{\phi}$ should be interpreted as effective density parameters, where $\Omega_{\phi}=x+y-f+g$. Using the slow-roll variable $N=\ln{a}$ and taking the derivatives with respect to $N$ one finds
\be\label{eq9}
f'=\frac{1}{H}\frac{dF}{dt}=\frac{1}{H}\left[\frac{\ddot{F}}{HF}-\frac{\dot{F}\dot{H}}{H^2F}-\frac{\dot{F}^2}{HF^2}\right]=\frac{\ddot{F}}{H^2F}-f\epsilon-f^2
\ee
\be\label{eq10}
k'=\frac{1}{H}\left[\frac{3F_1\dot{\phi}^2}{F}+\frac{6F_1\dot{\phi}\ddot{\phi}}{F}-\frac{3F_1\dot{\phi}^2\dot{F}}{F^2}\right]=\frac{3}{HF}\frac{d}{dt}(F_1\dot{\phi}^2)-kf
\ee
where " $'$ " means the derivative with respect to $N$. From the Eq. (\ref{eq4}) and using (\ref{eq9}) and (\ref{eq10}) follows
\be\label{eq11}
2\epsilon(1-\frac{1}{3}k)=-6x-(f'+f\epsilon+f^2)+f-2k+\frac{2}{3}(k'+kf)-3(1+w_m)\Omega_m.
\ee
Replacing the matter density parameter $\Omega_m$ from Eq. (\ref{eq8}) into Eq. (\ref{eq11}) leads to the equation
\be\label{eq11a}
2\epsilon(1-\frac{1}{3}k)=-6x-(f'+f\epsilon+f^2)+f-2k+\frac{2}{3}(k'+kf)-3(1+w_m)(1-x-y+f-k).
\ee
And taking the derivative w.r.t. $N$ for $x$ and $y$ from (\ref{eq7}), it is obtained
\be\label{eq12}
x'=\frac{1}{H}\left[\frac{\dot{\phi}\ddot{\phi}}{3H^2F}-\frac{\dot{H}\dot{\phi}^2}{3H^3F}-\frac{\dot{\phi}^2\dot{F}}{6H^2F^2}\right]=\frac{\dot{\phi}\ddot{\phi}}{3H^3F}-2x\epsilon -x f
\ee
\be\label{eq13}
y'=\frac{1}{H}\left[\frac{\dot{V}}{3H^2F}-\frac{2V\dot{H}}{3H^3F}-\frac{V\dot{F}}{3H^2F^2}\right]=\frac{\dot{V}}{3H^3F}-2y\epsilon-y f
\ee
Multiplying the equation of motion (\ref{eq5}) by $\dot{\phi}$ and using the product $\dot{\phi}\ddot{\phi}$ from (\ref{eq12}) one finds
\be\label{eq14}
x'+2x\epsilon+xf+6x+y'+2y\epsilon+yf-(2+\epsilon)f+\frac{1}{3}(k'+kf)+\frac{2}{3}(3+2\epsilon)k=0
\ee
In order to deal with the derivative of the potential and to complete the autonomous system we define the three parameters $b$, $c$ and $d$ that characterize the main properties of the model, as follows
\be\label{eq15}
b=\frac{1}{dF/d\phi}\frac{d^2F}{d\phi^2}\phi,\;\;\; c=\frac{1}{V}\frac{dV}{d\phi}\phi,\;\;\; d=\frac{1}{F_1}\frac{d F_1}{d\phi}\phi
\ee
These parameters are related to the potential and the couplings, and in what follows we restrict the model to the case when the parameters b, c and d are constant, which imply restrictions on the functional form of the couplings and potential. 
Additionally we introduce the new dynamical variable $\Gamma$ defined as
\be\label{eq16}
\Gamma=\frac{1}{F}\frac{dF}{\phi}\phi
\ee
using the constant parameters $b$, $c$, $d$ and the variable $\Gamma$, we can simplify the dynamical equations for the  variables  $y$, $f$, $k$ and $\Gamma$, reducing them to 
\be\label{eq17}
y'=\frac{c}{\Gamma}f y-2y\epsilon-y f
\ee
\be\label{eq18}
f'=\frac{b}{\Gamma}f^2+\frac{1}{2}f\frac{x'}{x}-\frac{1}{2}f^2
\ee
\be\label{eq19}
k'=2\epsilon k+\frac{d}{\Gamma}k f+k\frac{x'}{x}
\ee
\be\label{eq20}
\Gamma'=b f+f-\Gamma f
\ee
The equations (\ref{eq11a}) and (\ref{eq14}) together with the equations (\ref{eq17})-(\ref{eq20}) form the autonomous system. 
\section{The critical points}
By solving the simultaneous system of equations  (\ref{eq11a}), (\ref{eq14}) and (\ref{eq17})-(\ref{eq20}) with respect to  $x',y',f',k',\Gamma'$ and $\epsilon$, one finds
\be\label{eq21}
\begin{aligned}
x'=&-\frac{1}{D}\Big[x (2 f (d k (2 - f + 2 k + 4 x) + 3 b f (f - 2 (k + x)) + 3 c (2 + f - 2 k) y) + \\
  & \Gamma (3 f^3 + f^2 (-8 k + 18 w) + 12 (k^2 (-1 + 3 w) + 3 x (1 - x + y + w (-1 + x + y)) + \\
     &    k (-1 - 8 x + 3 y + 3 w (-1 + 2 x + y))) + 2 f (2 k^2 + k (5 - 27 w - 2 x) - \\
     &    3 (1 - 7 x + 3 y + 3 w (-1 + 3 x + y)))))\Big],
\end{aligned}
\ee
\be\label{eq22}
\begin{aligned}
y'=& \frac{1}{D}\Big[y (\Gamma (-3 f^3 + 4 f^2 (3 + 2 k) - 4 f (k^2 + (6 - 9 w) x - k (-6 + 3 w + x)) - \\
     & 12 (k + 3 x) (-1 + k (-1 + w) - x + y + w (-1 + x + y))) + f (4 b f (k + 3 x) - \\ &2 d k (f + 4 x) + 
     c (3 f^2 - 2 f (4 k + 3 y) + 4 (k + k^2 + 3 x - k x + 2 k y))))\Big],
\end{aligned}
\ee
\be\label{eq23}
\begin{aligned}
f'=&\frac{1}{D}\Big[f (f (2 b (2 k^2 + 3 (2 + f) x - k (-2 + f + 2 x)) + d k (f - 2 (1 + k + 2 x)) -  \\ & 3 c (2 + f - 2 k) y) +
   \Gamma (-3 f^3 + f^2 (8 k - 9 w) - 6 (k^2 (-1 + 3 w) + \\ &3 x (1 - x + y + w (-1 + x + y)) +
       k (-1 - 8 x + 3 y + 3 w (-1 + 2 x + y))) + \\ &f (-4 k^2 + k (-7 + 27 w + 4 x) + 
        3 (1 - 9 x + 3 y + 3 w (-1 + 3 x + y)))))\Big],
 \end{aligned}
\ee
\be\label{eq24}
\begin{aligned}
k'=&-\frac{1}{D}\Big[k (f (2 b f (3 f - 4 k) + d (-3 f^2 + 4 f k + 4 (-3 + k) x) - 4 c (-3 + k) y) + \\
  & \Gamma (3 f^3 + f^2 (12 - 8 k + 18 w) + 24 (k^2 w + 3 x + k (-2 x + y + w (-1 + x + y))) + \\
   &   2 f (2 k^2 - k (5 + 21 w + 2 x) - 3 (1 - 5 x + 3 y + 3 w (-1 + x + y)))))\Big],
\end{aligned}
\ee
\be\label{eq25}
\Gamma'=\left(1-\Gamma+b\right)f,
\ee
\be\label{eq26}
\begin{aligned}
\epsilon=&\frac{1}{D}\Big[f (-2 b f (k + 3 x) + d k (f + 4 x) + c (3 f - 4 k) y) - 
2 \Gamma (3 f^2 + \\ &f (-5 k + 3 k w - 3 x + 9 w x) - 
    3 (k + 3 x) (-1 + k (-1 + w) \\ &- x + y + w (-1 + x + y)))\Big],
\end{aligned}
\ee
where
$$D=\Gamma \left(3 f^2 - 8 f k + 4 (k + k^2 + 3 x - k x)\right).$$
The effective equation of state is given by $w_{eff}=-1-2\epsilon/3$. In order to find the critical points we need to solve the system of equations $x'=0$, $y'=0$, $f'=0$, $k'=0$, $\Gamma'=0$. To specify the model we need to define the scalar field dependence of the potential $V$ and the couplings $F$ and $F_1$. To this end, we use the fact that the parameters $b$, $c$ and $d$ are constants. \\
\noindent {\bf  1. Power-law couplings and potential}\\
In this case we consider that the parameters $b$, $c$ and $d$ are constants. In fact from (\ref{eq15}) follows the power-law behavior
\be\label{eq27}
h(\phi)\propto \phi^{b+1},\;\;\; V(\phi)\propto \phi^c,\;\;\; F_1(\phi)\propto \phi^{d}
\ee
where we used $F(\phi)=1/\kappa^2-h(\phi)$. The critical points for the system satisfy the equations 
\be\label{eq27a}
x'=0,\;\; y'=0,\;\; f'=0,\;\; k'=0\;\;, \Gamma'=0
\ee
 where the stability of the fixed points is determined by evaluating the eigenvalues of the Hessian matrix, associated with the system, at the critical points . In table 1 we arrange the critical points resulting from the solution of (\ref{eq27a}) with their main  associated quantities., and in table 2 we give the eigenvalues for each critical point.\\
\begin{table}[ht]
\caption{The critical points and some cosmological parameters for the model (\ref{eq1}) with the couplings and potential given in (\ref{eq27}).} 
\scriptsize
\centering 
\begin{tabular}{c c c c c c c c} 
\hline\hline \\
 & $x$ & $y$ & $f$ & $k$ & $\Gamma$ & $w_{eff}$  & $\Omega_\phi$ \\ [1ex] 
\hline\\ 
\bf{A1} & 1 & 0 & 0 & 0 & $1+b$ & 1  & 1\\ [1ex]
\hline\\
\bf{A2} & 0 & 0 & 0 & 1 & $\Gamma$ & $-1/3$  & 1 \\ [1ex]
\hline\\
\bf{A3} & $-\frac{1}{2}$ & 0 & 0 & $3/2$ & $1+b$ & $-1$  & 1 \\ [1ex]
\hline\\
\bf{A4} & $0$ & 0 & $-1$ & $0$ & $1+b$ & $1/3$  & 1 \\ [1ex]
\hline\\
\bf{A5} & $0$ & $\frac{{5 + 5b - c}}{{1 + b + c}}$ & $\frac{{4 + 4b - 2c}}{{1 + b + c}}$ & $0$ & $1+b$ & $-1+\frac{{2(1+b -c)(2+2b - c) }}{{3(1 + b)(1 + b + c)}}$  & 1 \\ [1ex]
\hline\\
\bf{A6} & $0$ & $0$ & $\frac{{1-3b+2d-f_1}}{{1 + 3b - d}}$ & $\frac{{2 + d-f_1}}{{1 + 3b - d}}$ & $1+b$ & $ - 1 + \frac{{(b - d-1)(3b-2d-1+f_1)}}{{3(b + 1)(3b-d+1)}}$  & 1 \\ [1ex]
\hline\\
\bf{A7} & $0$ & $0$ & $\frac{{1-3b+2d+f_1}}{{1 + 3b - d}}$ & $\frac{{2 + d+f_1}}{{1 + 3b - d}}$ & $1+b$ & $ - 1 +\frac{{(b - d-1)(3b-2d-1-f_1)}}{{3(b + 1)(3b-d+1)}}$ & $1$ \\ [1ex]
\hline\\
\bf{A8} & $0$ & $0$ & $1-3w_m$ & $0$ & $1+b$ & $ 1/3$ & $-1+3w_m$  \\ [1ex]
\hline\\
\end{tabular}
\label{ptoscriticos1} 
\end{table}

where $f_1$ represents the expression:

\[\begin{gathered}
 f_1=\sqrt{4d(d+4)-27b(b+2)-11}\end{gathered} \]
\clearpage

\begin{table}[ht]
\caption{The eigenvalues corresponding to the critical points of Table 1.} 
\scriptsize
\centering 
\begin{tabular}{c c c c c c} 
\hline\hline \\
 & $\lambda_1$ & $\lambda_2$ & $\lambda_3$ & $\lambda_4$ & $\lambda_5$  \\ [1ex] 
\hline\\ 
\bf{A1} & $-6$ & $6$ & 0 & 0 & $3(1-w_m)$\\ [1ex]
\hline\\
\bf{A2} & 3 & 3 & 3/2 & 0 & $-3w_m$ \\ [1ex]
\hline\\
\bf{A3} & $0$ & 0 & 0 & $-3$ & $-3(1+w_m)$ \\ [1ex]
\hline\\
\bf{A4} & $\frac{{b - 1}}{{b + 1}}$ & $1$ & $\frac{{5 + 5b - c}}{{1 + b}}$ & $ - \frac{{5 + 3b + d}}{{1 + b}}$ & $2-3w_m$ \\ [1ex]
\hline\\
\bf{A5} & $\frac{{2(1 - b)(2 + 2b - c)}}{{(1 + b)(1 + b + c)}}$ & $\frac{{2(c - 2 - 2b)}}{{1 + b + c}}$ & $\frac{{c - 5 - 5b}}{{1 + b}}$ & $\frac{{2(2 + 2b - c)(c + d  - 2b)}}{{(1 + b)(1 + b + c)}}$ & $-\frac{{(3 +6b+3{b^2}+7c+7bc-2{c^2}+3(1+b)(1+b+c) {w_m}) }}{{( 1 + b)(1 + b + c)}}$ \\ [1ex]
\hline\\
\bf{A6} & $ \frac{{1}}{{2}}$ & $ \frac{{d}}{{2(2+ d)}}$ & $\frac{{2-c+d}}{{2(2+ d)}}$ & $ - \frac{{5}}{{2-3{w_m}}}$ & $-\frac{{5}}{{2}}$ \\ [1ex]
\hline\\
\bf{A7} & $ \frac{{1}}{{2}}$ & $ \frac{{d}}{{2(2+ d)}}$ & $\frac{{2-c+d}}{{2(2+ d)}}$ & $ - \frac{{5}}{{2-3{w_m}}}$ & $-\frac{{5}}{{2}}$\\ [1ex]
\hline\\
\bf{A8} & $ \frac{{(b-1)(-1+3{w_m})}}{{1+b}}$ & $ -1+3{w_m}$ & $-2+3{w_m}$ & $  \frac{{3+c+3{w_m}(1-c)+3b(1+{w_m})}}{{1+b}}$ & $\frac{{d(1-3{w_m})-3(1+{w_m})-b(5-3{w_m})}}{{1+b}}$\\ [1ex]
\hline\hline 
\end{tabular}
\label{autovalores1} 
\end{table}

\noindent From Table 1 we can see that the critical point {\bf A1} is dominated by the kinetic energy of the scalar field ($\Omega_{\phi}=1$), with $w_{eff}=1$ corresponding to "stiff" matter, and 
is unstable critical point that could describe early time dominance of the scalar field.\\
The critical point {\bf A2} which is dominated by the kinetic coupling of the scalar field is unstable and gives an effective EoS that mimics dust-like matter.\\
The fixed point {\bf A3} is dominated by the scalar field and is a de Sitter solution with $w_{eff}=-1$. The negative sign of $x$ indicates phantom behavior and the eigenvalues indicate that at least the point is saddle. The three zero eigenvalues make difficult to analyze the stability, but since the rest of the eigenvalues are negative, the point is saddle.
This solution could correspond to an unstable inflationary phase which evolves towards a matter or dark energy dominated phase.\\
The point {\bf A4} is controlled by the non-minimal coupling and gives a solution that leads to an equation of state corresponding to radiation $w_{eff}=1/3$. At this critical point the potential and the kinetic coupling are absent and is a saddle point, depending on the values of the parameters $b,c,d$ and $w_m$. Thus for instance, if $-1<b<1$, $c>5(1+b)$, $d>-5-3b$ and $w>2/3$ all the eigenvalues except one are negative. For background radiation ($w_m=1/3$) or dust matter ($w_m=0$) three of the eigenvalues might take negative values. In the case of background matter given by radiation, this critical point presents a scaling behavior. At this point, despite the presence of the background matter in form of radiation or dust, the universe becomes radiation dominated, but due to the saddle character, this point could describe a transient phase of radiation dominated universe.\\ 
The critical point {\bf A5} is dominated by the potential and the non-minimal coupling with 
\be\label{eq28}
w_{eff}=-1+\frac{2 (1 + b - c) (2 + 2 b - c)}{3 (1 + b) (1 + b + c)},
\ee
and $\Omega_{\phi}=1$. The effective EoS describes different regimes depending on the parameters $b,c,d$ associated with the non-minimal coupling, the potential and the kinetic coupling. Note that for the scalar field dominated universe the effective EoS $w_{eff}$ and the dark energy EoS $w_{DE}$ take the same value. From (\ref{eq28}) follows that in the case $c=1+b$ we obtain the de Sitter solution with $w_{eff}=-1$, with eigenvalues given by $$\left[\frac{1 - b}{1 + b}, -1, -4, \frac{-b + d}{1 + b}, -4-3w_m\right].$$This solution is a stable fixed point for any type of matter with $0\le w_m\le1$, whenever $b>1$ and $d<b$ or $b<-1$ and $d<b$. The quadratic potential and the standard non-minimal coupling, corresponding respectively to $c=2$ ($V\propto \phi^2$) and $b=1$ ($h(\phi)\propto \phi^2$), lead to de Sitter solution, but in this case the eigenvalues are  $[0,-1,-4,\frac{1}{2}(-1+d),-4-3w_m]$ and the solution is marginally stable since four eigenvalues are negative (whenever $d<1$). The Higgs-type potential, $V\propto \phi^4$, corresponding to $c=4$ with non-minimal coupling $h\propto \phi^4$ ($b=3$), leads to de Sitter stable solution whenever $d<3$. The cubic non-minimal coupling, $h\propto \phi^3$, and cubic potential $V\propto \phi^3$, also give stable de Sitter solution with eigenvalues $[-(1/3), -1, -4, 1/3 (-2 + d), -4]$, for any $d<2$. The de Sitter solution can also be obtained for $c=2+2b$ with the eigenvalues $[0,0,-3,0,-3(1+w)]$, which contain three zeros, making difficult the stability analysis by the centre manifold method. 
We can also consider values for the effective EoS in the region of quintessence ($w_{eff}>-1$), or in the phantom region ($w_{eff}<-1$) , which are consistent with observations for $w_m$ in the interval $0\le w_m\le 1$. The conditions for the existence of stable quintessence fixed point are $b<-1$, $1+b< c< (3-\sqrt{10})(1+b)$ and $d> 1+2b-c$ or $b>1$, $(3-\sqrt{10})(1+b)< c< 1+b$ and $d<1+2b-c$. Thus, $b=4, c=4, d=1$, give a stable critical point with eigenvalues $[-4/5, -4/3, -21/5, -16/15, -61/15]$ and  $w_{eff}\approx -0.91$. The conditions for the existence of stable phantom solutions are $b<-1$, $2+2b<c< 1+b$ and $d>1+2b-c$ or $b>1$, $1+b< c<2+2b$ and $d<1+2b-c$. The parameters $b=2, c=4, d=0$ give a stable phantom solution with $w_{eff}\approx -1.06$ and eigenvalues  $[-4/21, -4/7, -11/3, -4/21, -79/21]$. \\
The coordinates of this fixed point give the behavior of the physical quantities related to the model. From $\epsilon$ defined in (\ref{eq7}) and the solution (\ref{eq26}) it is found
\be\label{eq29}
H=\frac{p}{t},\;\;\; p=\frac{(1+b)(1+b+c)}{(1+b-c)(2+2b-c)}
\ee
which gives 
\be\label{eq30}
a(t)=a_0t^p,\;\;\;  (p>0),\;\;\; \text{and}\;\;\; a(t)=\frac{a_0}{(t_c-t)^{|p|}},\;\;\; (p<0).
\ee
The last solutions leads to the known Big Rip singularity characteristic of the phantom power-law expansion. To find the scalar field we use the dynamical variables $f$ and $\Gamma$ defined in (\ref{eq7}) and (\ref{eq16}) taking into account their values at {\bf A5}
\be\label{eq31}
\frac{\dot{\phi}}{H\phi}=\frac{4 + 4 b - 2 c}{(1+b)(1 + b + c)}
\ee
which gives after integration gives
\be\label{eq32}
\phi=\phi_0 t^{\frac{2}{1+b-c}}\;\;\;  ({\it quintessence}),\;\;\; \phi=\phi_0 (t_c-t)^{\frac{2}{1+b-c}}\;\;\; ({\it phantom},\; p<0)
\ee
with these solutions, the asymptotic value $x\rightarrow 0$ at $t\rightarrow \infty$ ($t\to t_c$) is obtained by using $H\propto t^{-1}$ ($H\propto (t_c-t)^{-1}$), $h(\phi)\propto \phi^{b+1}$ and from (\ref{eq32}) $\phi\propto t^{\frac{2}{1+b-c}}$ ($\phi\propto (t_c-t)^{\frac{2}{1+b-c}}$)
\be\label{eq33}
x=\frac{\dot{\phi}^2}{6H^2 F}\propto \frac{t^{\frac{4}{1+b-c}}}{\kappa^{-2}-\xi t^{\frac{2(b+1)}{1+b-c}}}\; (p>0),\; x\propto \frac{(t_c-t)^{\frac{4}{1+b-c}}}{\kappa^{-2}-\xi (t_c-t)^{\frac{2(b+1)}{1+b-c}}}\; (p<0)
\ee
The inequalities $b>1$ and $c<1+b$ or $b<-1$ and $c>1+b$ lead to 
\be\label{eq33a}
\lim_{t\to \infty}x=0.
\ee
Thus, the conditions for stable quintessence solution satisfy this limit. And the conditions, $b<-1$ and $c<1+b$ or $b>1$ and $c>1+b$, lead to the limits
\be\label{eq33b}
\lim_{t\to t_c}x=0.
\ee
These conditions are satisfied by stable phantom solutions.\\
From the expression for $\Gamma$ (assuming $h(\phi)=\xi \phi^{b+1}$)
\be\label{eq34}
\Gamma=\frac{F'\phi}{F}=\frac{-\xi (b+1)\phi^{b+1}}{\kappa^{-2}-\xi\phi^{b+1}},
\ee
the limit
\be\label{eq35}
\lim_{t\to \infty}\Gamma=b+1,
\ee
for stable quintessence solutions takes place in the two cases: $b>1$ and $c<1+b$ or $b<-1$ and $c>1+b$. In the first case $\lim_{t\to \infty}\phi=\infty$, and in the second case $\lim_{t\to \infty}\phi=0$. For stable phantom solutions, the limit
\be\label{eq35a}
\lim_{t\to t_c}\Gamma=b+1,
\ee
takes place in two cases: $b<-1$ and $c<1+b$, where according to Eq. (\ref{eq32}), $\lim_{t\to t_c}\phi=0$, or $b>1$ and $c>1+b$, where $\lim_{t\to t_c}\phi=\infty$.
Concerning the effective Newtonian coupling, as follows from the definition $\kappa_{eff}^2=F(\phi)^{-1}$
\be\label{eq34a}
G_{eff}=\frac{G}{1-8\pi G \xi \phi^{b+1}},
\ee
we see that the restrictions on quintessence solutions lead to vanishing effective Newtonian coupling at $t\rightarrow \infty$, which also takes place for the phantom solutions, where at $t\to t_c$, $G_{eff}\to 0$,  indicating that the gravity reaches an asymptotic freedom regime (see \cite{ritis}) as the universe evolves towards the Big Rip singularity.\\
According to the EoS (\ref{eq28}), the de Sitter solution takes place for $c=1+b$, where the Hubble parameter becomes constant and the universe expands exponentially  
\be\label{eq36}
H=const=H_0, \;\;\; a(t)=a_0 e^{H_0(t-t_0)}
\ee
The scalar field can be found from the relation $f/\Gamma$ at the fixed point 
\be\label{eq37}
\frac{\dot{\phi}}{H\phi}=\frac{1}{1+b}.
\ee
where we have replaced $c=1+b$. Integrating this equation gives
\be\label{eq38}
\phi(t)=\phi_0 e^{\frac{H_0}{1+b}(t-t_0)}. 
\ee
Taking into account that the de Sitter solution is stable in the cases $b>1$ ($d<b$) and $b<-1$ ($d<b$), then the scalar field takes the asymptotic values
$$\lim_{t\to \infty}\phi=\infty \;\;\;\; {\it for}\;\;\;\;  b>1$$ and
$$\lim_{t\to \infty}\phi=0 \;\;\;\; {\it for}\;\;\;\; b<-1$$
To find the constant Hubble parameter in this case, we consider the critical value of the $y$-coordinate given by $\frac{5+5b-c}{1+b+c}=2$ and the definition of the variable $y$ given by the Eq. (\ref{eq7})
\be\label{eq35a}
y=\frac{V}{3H^2F}=\frac{V_0\phi^{b+1}}{3H_0^2(\kappa^{-2}-\xi\phi^{b+1})}=\frac{V_0\phi_0^{b+1}e^{H_0(t-t_0)}}{3H_0^2(\kappa^{-2}-\xi\phi_0^{b+1}e^{H_0(t-t_0)})}
\ee   
thus, according to this equation, the critical value of the $y$-coordinate (i.e. $y=2$) can be reached at $t\to \infty$, independently of the parameter $b$ since the power $b+1$ cancels with the denominator in the expression for the scalar field (\ref{eq38}) . Thus, we find the Hubble parameter as
\be\label{eq35b}
H_0^2=-\frac{V_0}{6\xi}.
\ee
Since we assume that the potential is positive (i.e. $V_0>0$), then this solution exists whenever $\xi<0$.
These results give us the behavior of $x$ from
\be\label{eq39}
\lim_{t\to \infty}x=\lim_{t\to \infty}\frac{\dot{\phi}^2}{6H^2F}
\ee
using  (\ref{eq36}) and (\ref{eq38}) for $H$ and $\phi$ we can see that 
\be\label{eq40}
x=\propto \frac{e^{\frac{2H_0}{1+b}(t-t_0)}}{\kappa^{-2}-\xi\phi_0^{b+1}e^{H_0(t-t_0)}}
\ee 
hence we find that for $b<-1$, when  $\lim_{t\to \infty}\phi=0$, we have $\lim_{t\to \infty} x=0$, and for the case
$b>1$, when $\lim_{t\to \infty}\phi=\infty$, then $\lim_{t\to \infty} x=0$, which are consistent with the solution (\ref{eq35b}). The coordinate $\Gamma$ from (\ref{eq34}) satisfies the limit $\lim_{t\to \infty}\Gamma=b+1$, for any $b$ as follows form the expression for the scalar field (\ref{eq38}). From the expression (\ref{eq34a}) for $G_{eff}$ we can conclude that when the fixed point becomes a de Sitter solution, the gravitational interaction reaches the asymptotic freedom regime, i.e. $G_{eff}\to 0$ at $t\rightarrow\infty$.\\
As seen from Table 1, the critical point {\bf A6} is dominated by the non-minimal and kinetic couplings, and from the expression for the effective EoS follows that the de Sitter solution takes place for  $b=d+1$. The expressions for the eigenvalues are too large to be displayed, and therefore we limit ourselves to the specific case of de Sitter solution, where we presented the real part of the eigenvalues.  As follows from the eigenvalues for the point {\bf A6}, the first eigenvalue prevents the stability of this point. 
From Table 1 for the point  {\bf A6} it can be seen that  $w_{eff}$ can not provide values in the interval $(-2,0)$, and takes only values in the interval $0\le w_{eff}\le 1$, which are interesting for early time cosmology where the behavior includes scaling solutions. These values take place for $b > -1$  and $d\ge 5 + 7 b$. So, the critical point {\bf A6} can not describe solutions with accelerated expansion.
Analyzing the stability in the relevant case $w_m=0$, and taking into account the above conditions for  $0\le w_{eff}\le 1$, it is found that the scaling solution with $w_{eff}=0$ is stable in the case $d=7b+5$, $b>1$ and $c<-5-5b$, and the scaling solution with $w_{eff}=1/3$ is unstable. \\
The critical point {\bf A7} is also dominated by the non-minimal and kinetic couplings, and presents the same characteristics and eigenvalues as the point {\bf A6}, leading to the same cosmological solutions. \\
To the fixed point {\bf A8} the matter and the non-minimal coupling contribute giving $w_{eff}=1/3$ with $\Omega_{\phi}=-1+3w_m$ and $\Omega_m=2-3w_m$. The positivity of the density parameters $\Omega_m$ and $\Omega_{\phi}$ impose the restriction $1/3\le w_m\le 2/3$, which excludes the pressureless dust matter. If the background matter consists of radiation ($w_m=1/3$), then the universe becomes radiation-dominated with $\Omega_{\phi}=0$ and $\Omega_m=1$, and the scaling solution mimics the radiation. At this saddle point with eigenvalues $[0,0,-1,4,-4]$ the system reaches the conformal invariance (given $w_m=1/3$) and can be considered as a transient phase of radiation dominated universe. In Fig. 1 we show the behavior of some trajectories around the critical point {\bf A5}, corresponding to de Sitter solution, for $b=1$, $c=2$.\\
\begin{center}
\includegraphics [scale=0.6]{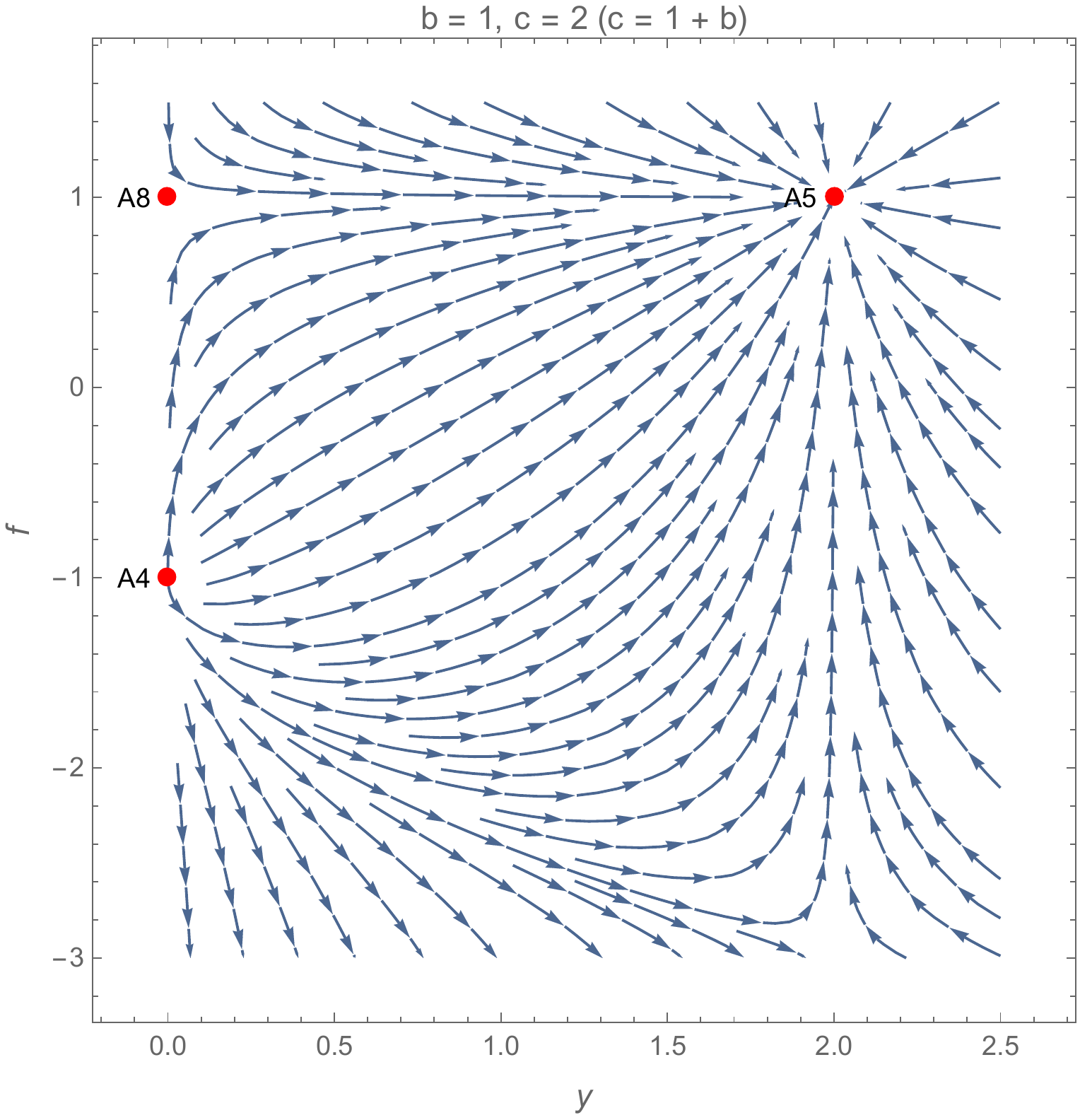}
\end{center} 
\begin{center}
{Fig. 1. \it The projection of the phase portrait of the model on the $yf$-plane for $b=1$ and $c=2$, assuming $w_m=0$. The attractor character of the de Sitter solution for the point {\bf A5} on the $yf$-plane is shown. The graphic shows trajectories evolving from the saddle points {\bf A4} and {\bf A8 } to the attractor {\bf A5}.}
\end{center}
The de Sitter solution shown in Fig. 1 corresponds to the standard non-minimal coupling $h\propto\phi^2$ and the quadratic potential $V\propto\phi^2$ ($b=1$, $c=2$), assuming $w_m=0$. The trajectories that converge to the de Sitter point {\bf A5}, evolve from the points {\bf A4} (saddle point, which attracts from the $k$-direction, corresponding to radiation dominated universe with $\Omega_{\phi}=1$) and {\bf A8}  (which is not physical since in this case $\Omega_m=2$).\\
There are two more critical points, namely
$$
\Big(0, \frac{(1 + b) (3 + c + 3 w_m - 3 c w_m + 3 b (1 + w_m))}{2 c^2}, 
 -\frac{3 (1+b)(1+w_m)}{c}, 0, 1 + b\Big)
 $$
 and
 $$
 \begin{aligned}
 &\Big(0,0,-\frac{3(1+b)(1+w_m)}{2b-d},\\& \frac{3(1+b)(1+w_m)((1-3w_m)d-3(1+w_m)+(3w_m-5)b)}{2(2b-d)(1+(1+2d)w_m+(1-3w_m)b)},1+b\Big)
 \end{aligned}
 $$
which are not of cosmological interest, since the density parameters fall out of the physical range. \\
\noindent {\bf  2. Exponential function for couplings and potential}\\
Here we impose the restrictions on the on the couplings and potential by redefining the  constant parameters $b$, $c$ and $d$ as
\be\label{eq43}
b=\frac{1}{dF/d\phi}\frac{d^2F}{d\phi^2},\;\;\; c=\frac{1}{V}\frac{dV}{d\phi},\;\;\; d=\frac{1}{F_1}\frac{d F_1}{d\phi}
\ee
with the new dynamical variable defined as
\be\label{eq43a}
\Gamma=\frac{F'}{F}
\ee
Integrating the equations (\ref{eq43}) with respect to the scalar field, one finds
\be\label{eq44}
h(\phi)\propto e^{b\phi},\;\;\; V(\phi)\propto e^{c\phi},\;\;\; F_1(\phi)\propto e^{d\phi}
\ee
where $b$, $c$ and $d$ are real numbers. The only equation of the autonomous system  (\ref{eq11a}), (\ref{eq14}) and (\ref{eq17})-(\ref{eq20}) that changes is the one related with the variable $\Gamma$ which reduces to
\be\label{eq45}
\Gamma'=\left(-\Gamma+b\right)f.
\ee
The critical points of the system are displayed in Table 3, with the respective eigenvalues given in Table 4. \\
\clearpage
\begin{table}[ht]
\caption{critical points and some cosmological parameters for the model  (\ref{eq1}) with the couplings and potential given in (\ref{eq43}).} 
\scriptsize
\centering 
\begin{tabular}{c c c c c c c c} 
\hline\hline \\
 & $x$ & $y$ & $f$ & $k$ & $\Gamma$ & $w_{eff}$  & $\Omega_\phi$ \\ [1ex] 
\hline\\ 
\bf{B1} & 1 & 0 & 0 & 0 & $b$ & 1  & 1\\ [1ex]
\hline\\
\bf{B2} & 0 & 0 & 0 & 1 & $\Gamma$ & $-1/3$  & 1 \\ [1ex]
\hline\\
\bf{B3} & $-\frac{1}{5}$ & 0 & 0 & $6/5$ & $b$ & $-1$  & 1 \\ [1ex]
\hline\\
\bf{B4} & $0$ & 0 & $-1$ & $0$ & $b$ & $1/3$  & 1 \\ [1ex]
\hline\\
\bf{B5} & $0$ & $\frac{{5b - c}}{{b + c}}$ & $\frac{{4b - 2c}}{{b + c}}$ & $0$ & $b$ & $ - 1 + \frac{{2(b - c)(2b - c)}}{{3b(b + c)}}$  & 1 \\ [1ex]
\hline\\
\bf{B6} & $0$ & $0$ & $\frac{{3b-2d+f_2}}{{d-3b}}$ & $\frac{{d-f_2}}{{3b - d}}$ & $b$ & $  - 1 + \frac{{(b - d)(3b-2d+f_2)}}{{3b(3b - d)}}$  & 1 \\ [1ex]
\hline\\
\bf{B7} & $0$ & $0$ & $\frac{{3b-2d-f_2}}{{d-3b}}$ & $\frac{{d+f_2}}{{3b - d}}$ & $b$ & $  - 1 + \frac{{(b - d)(3b-2d+f_2)}}{{3b(3b - d)}}$  & 1 \\ [1ex]
\hline\\
\bf{B8} & $0$ & $0$ & $1-3w_m$ & $0$ & $b$ & $ 1/3$ & $-1+3w_m$  \\ [1ex]
\hline\hline 
\end{tabular}
\label{ptoscriticos1} 
\end{table}

where $f_2$ represents the expression:
\[\begin{gathered}
 f_2=\sqrt{4d^2-27b^2} \end{gathered} \]
\clearpage

\begin{table}[ht]
\caption{The eigenvalues for the critical points of Table 3.} 
\scriptsize
\centering 
\begin{tabular}{c c c c c c} 
\hline\hline \\
 & $\lambda_1$ & $\lambda_2$ & $\lambda_3$ & $\lambda_4$ & $\lambda_5$  \\ [1ex] 
\hline\\ 
\bf{B1} & $0$ & $0$ & -6 & 6 & $3(1-w_m)$\\ [1ex]
\hline\\
\bf{B2} & 0 & 3/2 & 3 & 3 & $-3w_m$ \\ [1ex]
\hline\\
\bf{B3} & $0$ & 0 & 0 & $-3$ & $-3(1+w_m)$ \\ [1ex]
\hline\\
\bf{B4} & $\frac{{1}}{{3b}}$ & $1$ & $\frac{{5b - c}}{{b}}$ & $ - \frac{{3b + d}}{{b}}$ & $2-3w_m$ \\ [1ex]
\hline\\
\bf{B5} & $\frac{{2(c - 2b)}}{{b + c}}$ & $-\frac{{b+c}}{{6b(2b - c}}$ & $\frac{{c - 5b}}{b}$ & $\frac{{2( 2b - c)(c + d - 2b)}}{{b(b + c)}}$ & $\frac{{ 2{c^2}- 3{b^2}(1+{w_m}) - bc(7+{w_m})}}{{b(b + c)}}$ \\ [1ex]
\hline\\
\bf{B6} & $ - \frac{{1}}{{12d}}$ & $ \frac{{1}}{{2}}$ & $\frac{{d-c}}{{2d}}$ & $- \frac{{5}}{{2}}$ & $-\frac{{5}}{{2}}$ \\ [1ex]
\hline\\
\bf{B7} & $ - \frac{{1}}{{12d}}$ & $ \frac{{1}}{{2}}$ & $\frac{{d-c}}{{2d}}$ & $- \frac{{5}}{{2}}$ & $-\frac{{5}}{{2}}$\\ [1ex]
\hline\\
\bf{B8} & $ - \frac{{1}}{{3b(1-3{w_m})}}$ & $ -1-3{w_m}$ & $-2-3{w_m}$ & $\frac{{c(1-3{w_m})+3b(1+{w_m})}}{{b}}$ & $\frac{{d(1-3{w_m})-b(5-3{w_m})}}{{b}}$ \\ [1ex]
\hline\hline 
\end{tabular}
\label{autovalores1} 
\end{table}

\noindent Note that the points {\bf  B1}, {\bf  B2}, {\bf  B3} and {\bf  B4} lead to the same asymptotic values of the physical parameters with the same stability properties. \\
This fixed point  {\bf B5} is dominated by the scalar field, specifically by the potential and non-minimal coupling, with $\Omega_{\phi}=1$. From the effective EoS follows the scaling solution with $w_{eff}=w_m$ if the parameter $c$ is restricted as 
\be\label{eq47}
c=\frac{1}{4}\left(9b+3bw_m-b\sqrt{73+78w_m+9w_m^2}\right).
\ee
Replacing this restriction for $c$ in the eigenvalues for {\bf B5} (see Table 4) we find that the scaling solution corresponding to this critical point is stable for $0\le w_m\le 1$ if the inequalities $b>0$ and $d<\frac{1}{4}(-b-3bw_m+\sqrt{73b^2+78b^2w_m+9b^2w_m^2})$ are satisfied. So, if we define the potential so that the potential parameter $c$ depends on the non-minimal coupling parameter $b$ and $w_m$ as given by the equation (\ref{eq47}), then the critical point is a scaling attractor if the above inequalities are satisfied. This result provides a cosmological scenario where the energy density of the scalar field behaves similarly to the background fluid in either the radiation or matter era, but with the dominance of the scalar field.\\
\noindent According to the effective EoS, given in Table 3, the de Sitter solution takes place for $c=b$ and $c=2b$. In the case $c=b$ the eigenvalues reduce to $[-1,-1,-4,-1+\frac{d}{b},-4-3w]$, indicating that the de Sitter solution is a stable node (attractor) for any type of matter with $0\le w_m \le 1$ and for $d<b$, and is a saddle point if $d>b$. As follows from the expression for the eigenvalues, the case $c=2b$ leads to zero and indeterminate eigenvalues and therefore can not be considered. 
On the other hand, the quintessence behavior ($w_{eff}>-1$) takes place for the restriction $\frac{2(b-c)(2b-c)}{3b(b+c)}>0$. To analyze the stability in this case, we limit ourselves to the relevant interval $0\le w_m\le 1$, and them according to the expression for the eigenvalues, the quintessence fixed point is an attractor if the inequalities $b>0$, $\frac{1}{4}(7+3w_m-
\sqrt{73+66w_m+9w_m^2})b<c<b$ and $d<2b-c$ are satisfied.
The fixed point describes phantom phase or super accelerated expansion in the case $\frac{2(b-c)(2b-c)}{3b(b+c)}<0$. This phase is stable if the parameters satisfy:  $b>0$, $b<c<2b$ and $d<2b-c$. In the quintessence and phantom phases the effective EoS $w_{eff}$ can be as close to $-1$ as we need, since the parameters $b$, $c$ and $d$ are real numbers. So, this new fixed point is very interesting cosmological solution since it can account for different regimes of the accelerating universe.\\
We can analyze the behavior of the physical quantities that follows from this critical point. 
According to (\ref{eq7}) and (\ref{eq26}) for $\epsilon$ and using the coordinates of the fixed point {\bf B5} one finds
\be\label{eq47a}
\frac{\dot{H}}{H}=-\frac{(b-c)(2b-c)}{b(b+c)},
\ee
which leads to the scale factor
\be\label{eq47b}
a=a_0 t^{\gamma},\;\;\; \gamma=\frac{b(b+c)}{(b-c)(2b-c)}.
\ee
For the phantom solution (negative power) one can write $a=a_0(t_c-t)^{-|\gamma|}$. By using the relation between the variables $f$ (Eq. (\ref{eq7})) and $\Gamma$ (Eq. (\ref{eq16})) at the fixed point, one finds equation for the scalar field 
\be\label{eq47c}
\frac{f}{\Gamma}\Big |_{B5}=\frac{4b-2c}{(b+c)b}=\frac{\dot{\phi}}{H\phi}.
\ee
which gives the scalar field solution
\be\label{eq47d}
\phi=\phi_0 t^{\frac{2}{b-c}}\;\;\;  ({\it quintessence}),\;\;\; \phi=\phi_0 (t_c-t)^{\frac{2}{b-c}}\;\;\; ({\it phantom},\; \gamma<0)
\ee
From (\ref{eq7}) for $x$ and using the above results we can write the behavior of $x$ at $t\rightarrow \infty$ as
\be\label{eq47e}
x\propto \frac{(t)^{\frac{4}{b-c}}}{\kappa^{-2}-\xi e^{b\phi_0 (t)^{\frac{2}{b-c}}}}\; (\gamma>0),\; x\propto \frac{(t_c-t)^{\frac{4}{b-c}}}{\kappa^{-2}-\xi e^{b\phi_0 (t_c-t)^{\frac{2}{b-c}}}}\; (\gamma<0)
\ee
where we used $h(\phi)=\xi e^{b\phi}$. For stable quintessence solutions, from this expression we find that $x\to 0$ at $t \to \infty$ in the cases $b>0$ and $b>c$. For stable phantom solution it is found that $x\to 0$ at $t \to t_c$ for $b>0$ and $b<c$. Concerning the asymptotic behavior of the $\Gamma$-coordinate, it is found
\be\label{eq47f}
\Gamma\Big |_{B5}=\lim_{t\to \infty}\left(\frac{-\xi b e^{b\phi}}{\kappa^{-2}-\xi e^{b\phi}}\right).
\ee
From which follows that, if $b>0$, $b>c$ then $\lim_{t\to \infty}\phi=\infty$ and  $\lim_{t\to \infty}\Gamma=b$, and 
if $b>0$, $b<c$ then $\lim_{t\to t_c}\phi=\infty$ and
$\lim_{t\to t_c}\Gamma=b$. 
For stable quintessence solutions ($b>0$ and $b>c$) the effective Newtonian coupling $G_{eff}\to 0$ and the system reaches the asymptotic freedom regime. There are also quintessence solutions in the cases $b<0$, $b<c$ and $b>0$, $c>2b$, where $\Gamma\to b$ in the strong coupling regime and the effective Newtonian coupling becomes constant. For stable phantom solutions ($b>0$ and $b<c$) the gravitational interaction also reaches the asymptotic freedom regime. There are phantom solutions in the case $b<0$, $b>c$ where $\Gamma\to b$ in the strong coupling regime and $G_{eff}$ tends to a constant value, and in the case $b>0$, $c<-b$ where $\Gamma\to b$ in the strong coupling regime and $G_{eff}\to 0$. \\
\noindent From the EoS (see {\bf B5} in Table 3) follows that the de Sitter solution takes place for $c=b$, with the Hubble parameter and scale factor given by   
\be\label{eq56a}
H=const=H_0, \;\;\; a(t)=a_0 e^{H_0(t-t_0)}
\ee
The relation $f/\Gamma$ at this point gives (see (\ref{eq47c}))
\be\label{eq56b}
\phi=\phi_0 e^{\frac{H_0}{b}(t-t_0)}
\ee
The region of stability of the de Sitter solution is defined by $d<b$, in which the scalar field can take the limits
$$\lim_{t\to \infty}\phi=\infty \;\;\;\; {\it for}\;\;\;\;  b>0$$ and
$$\lim_{t\to \infty}\phi=0 \;\;\;\; {\it for}\;\;\;\; b<0$$
To find the value of the Hubble parameter, we consider the critical value of the $y$-coordinate given by $\frac{5b-c}{b+c}=2$ and the definition of the variable $y$ given by the Eq. (\ref{eq7}) (where we used $c=b$)
\be\label{eq56c}
2=\frac{V}{3H^2F}\Big |_{B5}=\lim_{t\to \infty}\frac{V_0 e^{b\phi}}{3H_0^2(\kappa^{-2}-\xi e^{b\phi})}
\ee   
Thus, taking into account the above limits for $\phi$, one finds that if $b>0$ then  $\lim_{t\to \infty}e^{b\phi}=\infty$ and the Hubble parameter takes the value
\be\label{eq56d}
H_0^2=-\frac{V_0}{6\xi}.
\ee
implying that $\xi<0$ for positive potential $V_0>0$. On the other hand, if $b<0$, then  $\lim_{t\to \infty}e^{b\phi}=1$, which leads to the Hubble parameter
\be\label{eq56e}
H_0^2=\frac{V_0}{6(\kappa^{-2}-\xi)}.
\ee
so, the solution exists whenever $\xi<0$ or $\xi<\kappa^{-2}$. Concerning the $x$-coordinate, which behaves as
\be\label{eq56f}
x\propto \frac{e^{\frac{2H_0}{b}(t-t_0)}}{\kappa^{-2}-\xi e^{b\phi_0 e^{\frac{H_0}{b}(t-t_0)}}},
\ee
we find that for any $b$ (independently of the sign),  $\lim_{t\to \infty} x=0$, which is consistent with the critical value. \\
After replacing the scalar field (\ref{eq56b}) in the expression (\ref{eq47f}) for the $\Gamma$-coordinate, one finds, for $b>0$ that $\lim_{t\to \infty}\Gamma=b$, and the effective Newtonian coupling vanishes at this limit, reaching the asymptotic freedom regime.
In the case $b<0$, at ${t\to \infty}$ we can consider the approximation of the strong coupling limit where $|\xi|>>\kappa^{-2}$, which leads to $\Gamma\to b$. In this limit and under the strong coupling approximation (with $\xi<0$) the de Sitter solution takes place with a constant effective Newtonian coupling ($G_{eff}\sim 1/(16\pi |\xi|)$).\\
As follows form the results in Table 3, the fixed point {\bf B6} exists for $-\frac{2|d|}{3\sqrt{3}}<b<\frac{2|d|}{3\sqrt{3}}$ and is dominated by the non-minimal and kinetic couplings, with $\Omega_{\phi}=1$, and the effective EoS given in Table 3. The de Sitter fixed point takes place in the case $b=d$, and in Table 4 we have shown the real part of the eigenvalues for the case of de Sitter solution with   $w_m=0$.
Note that there is one positive eigenvalue that spoils the stability of the de Sitter fixed point. Despite the fact that there is a de Sitter solution, $w_{eff}$ presents discontinuities at $b=0$ and $d=3b$ and is real only if the inequalities $d<-\frac{3\sqrt{3}}{2}|b|$ or $d>\frac{3\sqrt{3}}{2}|b|$ are satisfied. Therefore $w_{eff}$ can not take interesting values (i.e. $<-1/3$) in the interval $(-1,0)$, except for values very close to $0$. The scalar field mimics the pressureless matter ($w_{eff}=0$) in the case $d=7b$, and this solution is stable under the conditions $b>0$ and $c<-5b$. The radiation dominated solution is obtained for $d=-3b$, but this solution is unstable. 
Considering the limit $d\to 3b$, the radiation dominated solution is also with eigenvalues
$$
\Big[\frac{1}{6b},-2,2(1+\frac{c}{b}),-\frac{1}{2}(4+3w_m-|3w_m-2|,-2-\frac{3}{2}w_m-\frac{1}{2}|3w_m-2|\Big],
$$
which is stable whenever $b<0$ and $c>-b$.\\
The phantom solutions are also possible but they are unstable (saddle) with values that deviate away from the expected values for late time cosmology.\\
The fixed point {\bf B7} leads to the same results as the point {\bf B6}.\\
The fixed point {\bf B8} gives the same results as the point {\bf A8}. In Fig. 2 we show the behavior of some trajectories around the critical point {\bf B5}, showing the attractor character of the phantom solution with $w_{eff}=-1.06$, for $b=1$, $c=1.29$.\\\\
\begin{center}
\includegraphics [scale=0.6]{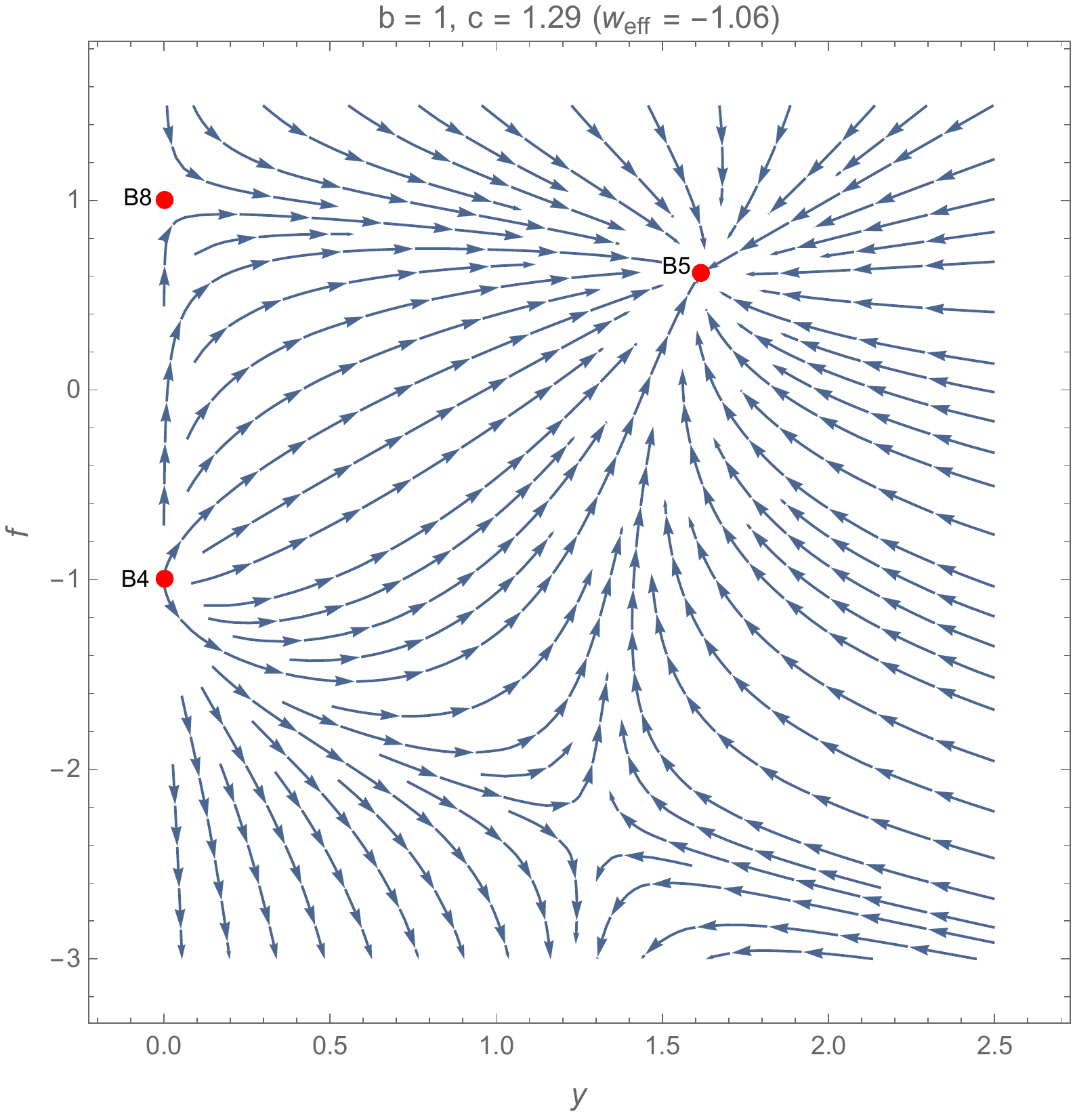}
\end{center} 
\begin{center}
{Fig. 2. \it The projection of the phase portrait of the model on the $yf$-plane for $b=1$, $c=1.29$ and $w_m=0$. The graphic shows that the phantom solution for the point {\bf B5}, with $w_{eff}=-1.06$, behaves as an attractor. The trajectories evolve from the points {\bf B4} and {\bf B8} to the phantom attractor {\bf B5}.}
\end{center}
There is a fixed point {\bf B9} with coordinates ($x=0$, $y=\frac{b(c-3cw_m+3b(1+w_m))}{2c^2}$, $f=-\frac{3b(1+w_m)}{c}$, $k=0$, $\Gamma=b$). The contribution of the scalar field and matter to this point are $\Omega_{\phi}=\frac{b(3b(1+w_m)+c(7+3w_m))}{2c^2}$ and $\Omega_m=\frac{2c^2-3b^2(1+w_m)-bc(7+3w_m)}{2c^2}$. The effective EoS is given by $w_{eff}=-1-\frac{(b-c)(1+w_m)}{c}$. The de Sitter solution takes place for $b=c$, but the corresponding density parameters fall out of the physical range. This is also the case for quintessence or phantom solutions.\\
To the last fixed point {\bf B9}, with coordinates  ($x=0$, $y=0$, $f=-\frac{3b(1+w_m)}{2b-d}$, $k=\frac{3b(1+w_m)(d-3d w_m-b(5-3w_m))}{2(2b-d)(2dw_m-b(1-3w_m))}$, $\Gamma=b$), contribute the kinetic coupling and matter with 
$\Omega_{\phi}=\frac{3b(3b-d)(1+w_m)^2}{2(2b-d)(-2dw_m-b(1-3w_m))}$ and $\Omega_{\phi}=\frac{4d^2w_m+bd(5-8w_m+3w_m^2)-b^2(13+6w_m+9w_m^2)}{2(2b-d)(-2dw_m-b(1-3w_m))}$. The effective EoS is given by $w_{eff}=-1+\frac{(b-d)(1+w_m)}{2b-d}$. The de Sitter solution exists but as in the previous case de density parameters take the values $\Omega_{\phi}=\frac{3(1+w_m)^2}{w_m-1}$ and $\Omega_m=\frac{4+5w_m+3w_m^2}{1-w_m}$, which are not physical for the cases $w_m=0$ and $w_m=1/3$ (i.e. do not satisfy $0\le \Omega_{\phi,m}\le 1$). Note however that if we set $b=0$, then $w_{eff}=w_m$ with $\Omega_{\phi}=0$ and $\Omega_m=1$, and in the case $w_m=0$ the eigenvalues become $[-\frac{1}{d},0,\frac{3(c+d)}{d},-\frac{3}{2},0]$ leading to a saddle point for $d>0$ and $c<-d$. For the matter component with $0<w_m\le 1$, the density parameters acquire values out of the physical region.\\
\section{Discussion}
In this work we studied some aspects of late-time cosmological dynamics for the scalar-tensor model with non-minimal coupling of the scalar field to curvature and non-minimal kinetic coupling to the Einstein tensor (see Eqs. (\ref{eq1}) and (\ref{eq1a})). For this study we considered two types of couplings and potential: power-law functions for the couplings and the potential, i.e., $h(\phi)\propto \phi^{b+1}$, $F_1(\phi)\propto \phi^{d+1}$ and $V(\phi)\propto \phi^c$, and exponential functions for the couplings and potential of the form $h(\phi)\propto e^{b\phi}$, $F_1(\phi)\propto e^{d\phi}$ and $V(\phi)\propto e^{c\phi}$. Of special interest are stable critical points leading to relevant late time cosmological scenarios, consistent with observations.\\ 
\noindent The presence of the kinetic coupling gives additional solutions with respect to the model of scalar field with non-minimal coupling that has been already considered in \cite{msami} (where power-law functions were considered). Of special relevance is the point {\bf A5} since it contains stable quintessence and phantom solutions besides the stable de Sitter solution. This point is dominated by the scalar field and the effective EoS at this critical point allows different viable late time cosmological solutions.   
The de Sitter solution takes place under the restriction $c=b+1$, and the stability is obtained for $b>1$ or $b<-1$ and $d<b$. The  particular case of $b=1$, which gives the standard non-minimal coupling $\xi \phi^2$, with $c=2$ ($V\propto \phi^2$), and $d=-2$ which gives dimensionless kinetic coupling constant, leads to de Sitter solution with marginal stability since 
the eigenvalues are $[0,-1,-4,-3/2,-4-3w_m]$. 
The Higgs-type potential ($V\propto \phi^4$) corresponding to $c=4$, with non-minimal coupling $h\propto \phi^4$ ($b=3$) together with $F_1=\chi \phi^{-2}$ ($d=-2$), leads to stable de Sitter expansion. Note the that the kinetic coupling of the form $F_1=\chi \phi^{-2}$  (where $\chi$ is dimensionless) is valid for all stable de Sitter solutions. It also has been shown that when the fixed point becomes a de Sitter solution, the gravitational interaction reaches the asymptotic freedom regime ($G_{eff}\to 0$).   
This point can also give stable solutions with equation of state in the region around $w_{eff}=-1$, with values above or below $-1$, consistent with current observational data. 
 The restrictions on stable quintessence and phantom solutions lead to vanishing effective Newtonian coupling at $t\to \infty$ and $t\to t_c$ respectively, indicating that the system reaches an asymptotic freedom regime. \\
The exponential functions for the couplings between the scalar field and curvature, which are typical of string-inspired gravity models, give rise to new stable quintessence and phantom solutions, including also de Sitter solutions. 
The critical point {\bf B5} contains a de Sitter solution for $b=c$, which is an attractor node for $b>d$ and saddle for $b<d$. The consistency with the coordinates of the fixed point in the case $b>0$ leads to the vanishing of the effective Newtonian coupling,
where the gravity reaches the asymptotic freedom regime. This behavior is different in the case $b<0$, where the effective Newtonian coupling tends to a constant value ($G_{eff}\sim 1/(16\pi|\xi|)$) at $t\to \infty$. For the quintessence and phantom solutions, the effective Newtonian coupling vanishes for $b>c$ at $t\to \infty$ for quintessence, and for $b<c$ at $t\to t_c$ for phantom solutions.\\
An interesting consequence of the exponential models is the existence of de Sitter solution where the effective Newtonian coupling becomes constant. The parameters $b$, $c$ and $d$ for the exponential functions can take real values, which allow to adjust the EoS of the dark energy to asymptotic values as close to $-1$ as required by observations. For the power-law functions these parameters were restricted to take integer values. Additionally, in all the above solutions the phantom scenario could be realized without introducing ghost degrees of freedom, which is quite attractive for a viable model of dark energy.
The present results show that the effect of the non-minimal and kinetic couplings can account for different accelerating regimes of the early and late time universe.

\section*{Acknowledgments}
This work was supported by Universidad del Valle under project CI 71074 and by COLCIENCIAS grant number 110671250405, DFJ acknowledges support from COLCIENCIAS, Colombia.


\end{document}